\documentclass[11pt,twoside]{article}
\usepackage{asp2004}
\usepackage{psfig}
\usepackage{epsf}
\usepackage{graphics}
\usepackage{lscape}
\begin{document}

\title{What can \emph{XMM-Newton\/} tell us about the spin periods of
Intermediate Polars?}
\author{P. A. Evans \&\ Coel Hellier}
\affil{Astrophysics Group, School of Chemistry and Physics, Keele
University, Keele, Staffordshire, ST5 5BG}

\begin{abstract} 
\emph{XMM-Newton}'s unprecedented combination of spectral resolution
and high throughput allows us to perform the best phase-resolved
X-ray analysis of intermediate polars to date. The Optical Monitor
gives optical/UV photometry simultaneously with the X-ray data.  We
present a comprehensive study of X-ray spin pulses in IPs, giving
spin-pulses and hardness ratios for every IP looked at with
\emph{XMM-Newton\/} to date.
\end{abstract}

\section{Introduction}
\label{sec:intro}

In an intermediate polar (IP) the white dwarf has a magnetic field
strong enough to affect the accretion flow, but not strong enough to
synchronise the white dwarf rotation with the binary period; the
magnetic dipole is also inclined to the spin axis of the star.
Generally accretion occurs from a magnetically truncated accretion
disc; at some radius the disc material threads to the white dwarf's
field lines and is channeled towards the gravitationally preferable
pole in large accretion curtains. Stand-off shocks form above each
pole, resulting in hard X-ray emission. As the white dwarf rotates,
our view of these regions changes and the position of the accretion
curtains relative to our line of sight also varies, giving rise to
spin-period modulations (e.g.\ Hellier, Cropper, \&\ Mason 1991).

To date eleven IPs have been observed with \emph{XMM-Newton\/}. In
this paper we present a compilation of X-ray spin-pulse profiles,
hardness ratios, and UV data from \emph{XMM}'s Optical Monitor (OM)
for these systems.

\section{AO Piscium and V1223 Sagittarii}
\label{sec:aov1223}

\begin{figure}
\begin{center}
\plottwo{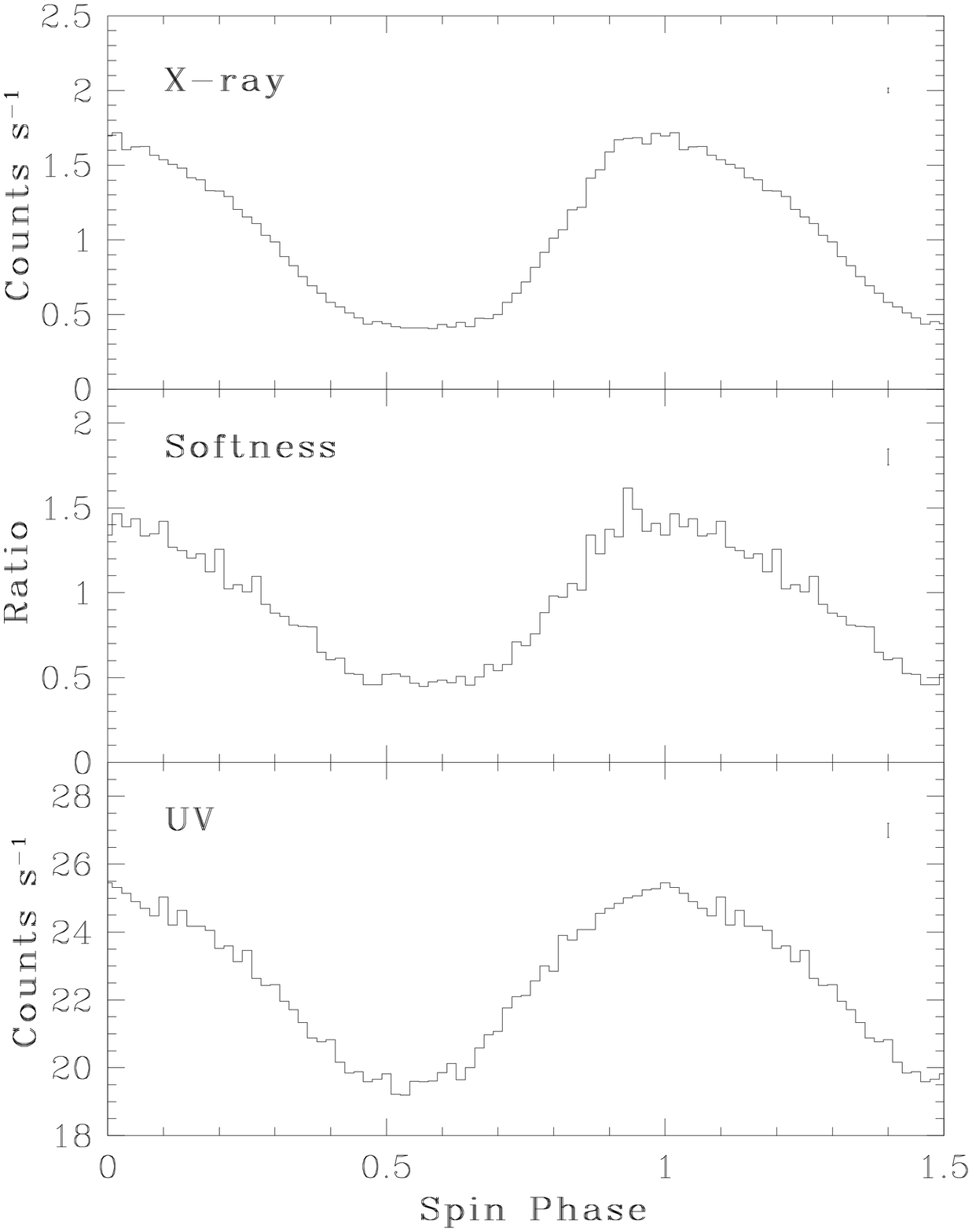}{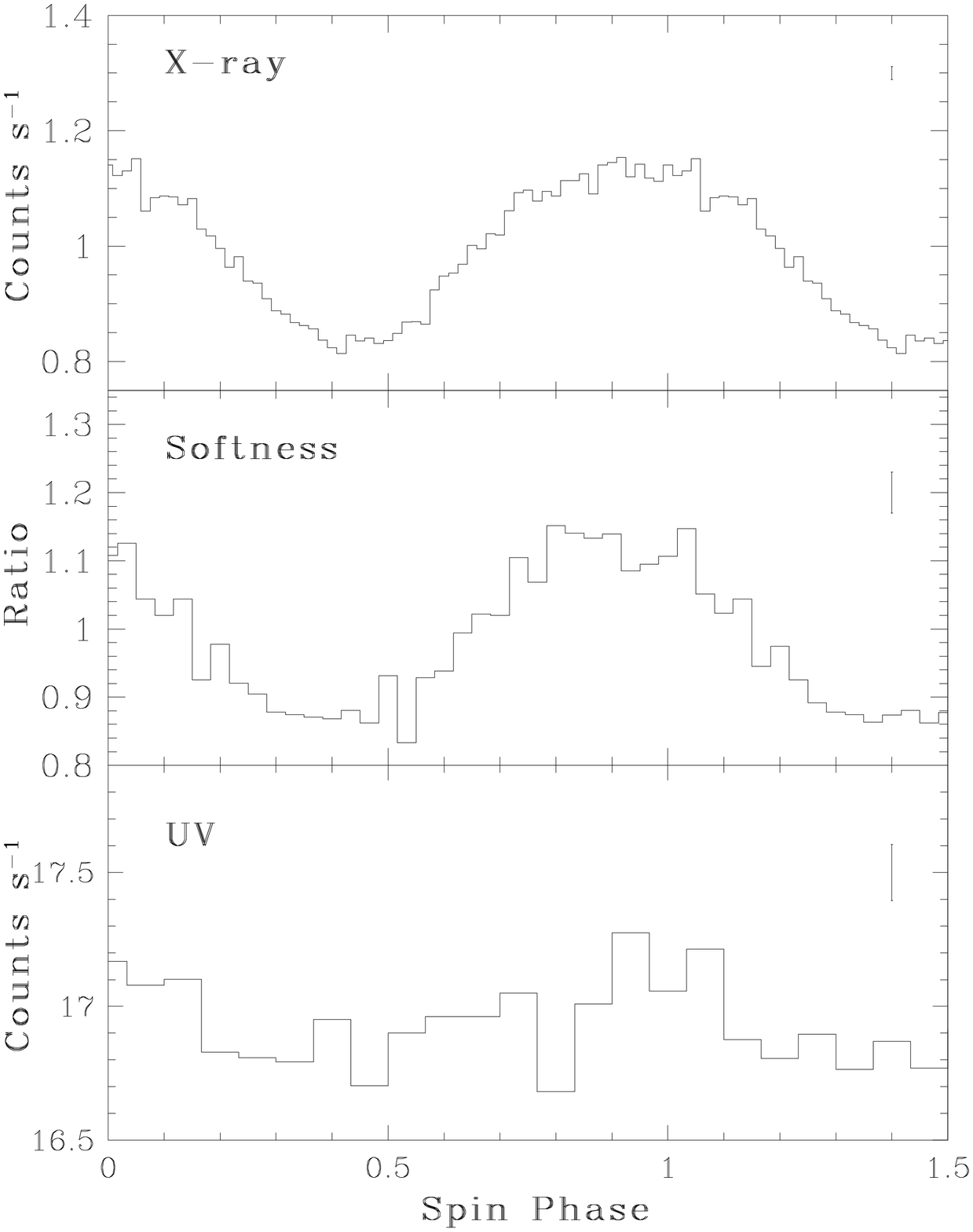}
\caption{The spin pulses of AO~Psc (left) and V1223~Sgr (right). We
show the 0.2--12 keV X-ray data, the (0.2--4)/(6--12) keV softness
ratio and (where available) the UV (2050--2450 \AA) data.}
\label{fig:ao}
\label{fig:v1223}
\end{center}
\end{figure}

AO Psc and V1223 Sgr show sinusoidal X-ray and UV modulations
(Fig.~\ref{fig:ao}). The fact that their spectra get harder at pulse
minimum suggests that the modulation is caused by absorption. These
are thus good examples of the accretion curtain model for X-ray
modulation in IPs (e.g.\ Hellier et al.\ 1991).

\section{HT Camelopardaris}
\label{sec:ht}

\begin{figure}
\begin{center}

\plotfiddle{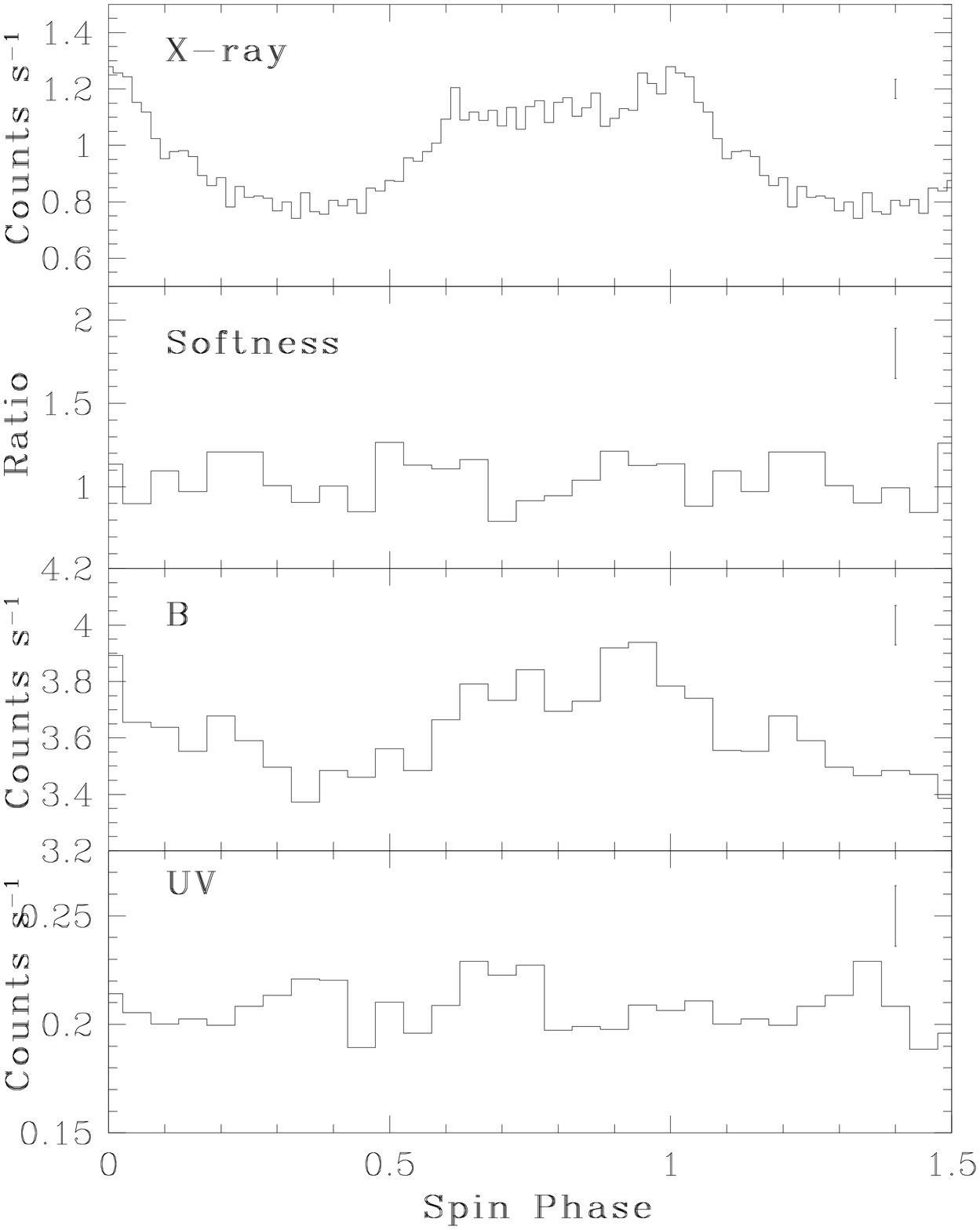}{8cm}{0}{32}{35}{-190}{-30}
\plotfiddle{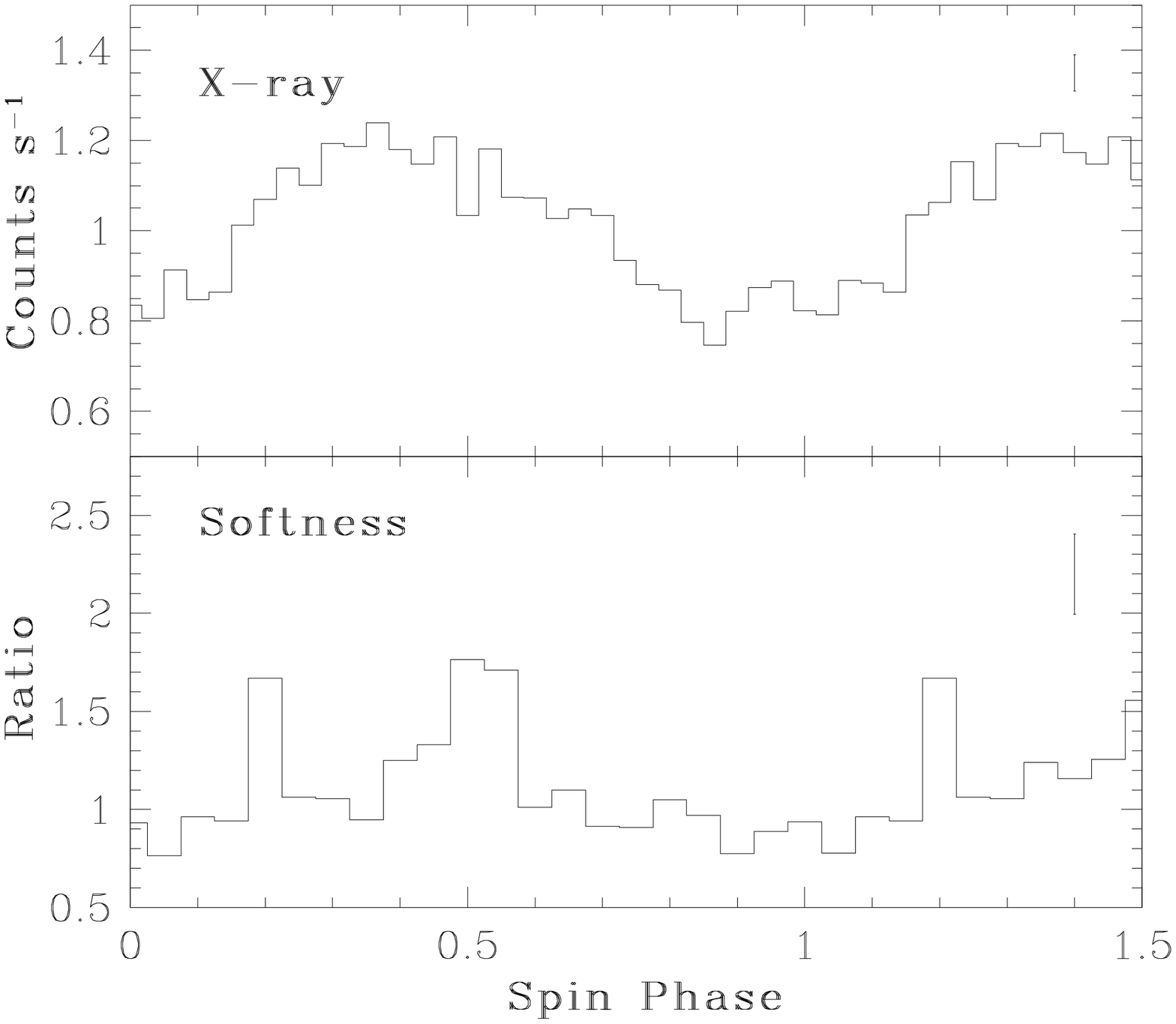}{-2cm}{0}{32}{35}{0}{-5}
\caption{As for Fig.~1, but for HT~Cam (left) and RX\,J1548.2$-$4528
(right). The UV band for HT~Cam is 1800--2250 \AA, and the $B$ band
(3900--4900 \AA) is also shown.}
\label{fig:ht}
\label{fig:rxs}
\end{center}
\end{figure}

HT~Cam's X-ray spin-pulse (Fig.~\ref{fig:ht}) appears sinusoidal with a
flattened maximum, and, judged from the softness ratio, appears to be energy
independent. No pulse is detected in the UV, although the $B$-band shows
sinusoidal variation. For more information, see de~Martino et al.\ (2004b).

\section{RX\,J1548.2$-$4528}
\label{sec:rxj}

The recently discovered IP RX\,J1548.2$-$4528 shows a sinusoidal
X-ray spin-pulse (Fig.~\ref{fig:rxs}). Haberl, Motch \&\ Zickgraf
(2002) analysed this observation, and suggested that while there may
be a small change in absorption with spin phase, the predominant
source of modulation is a change in the visibility of the emission.

\section{EX Hydrae}
\label{sec:ex}

\begin{figure}
\begin{center}
\plottwo{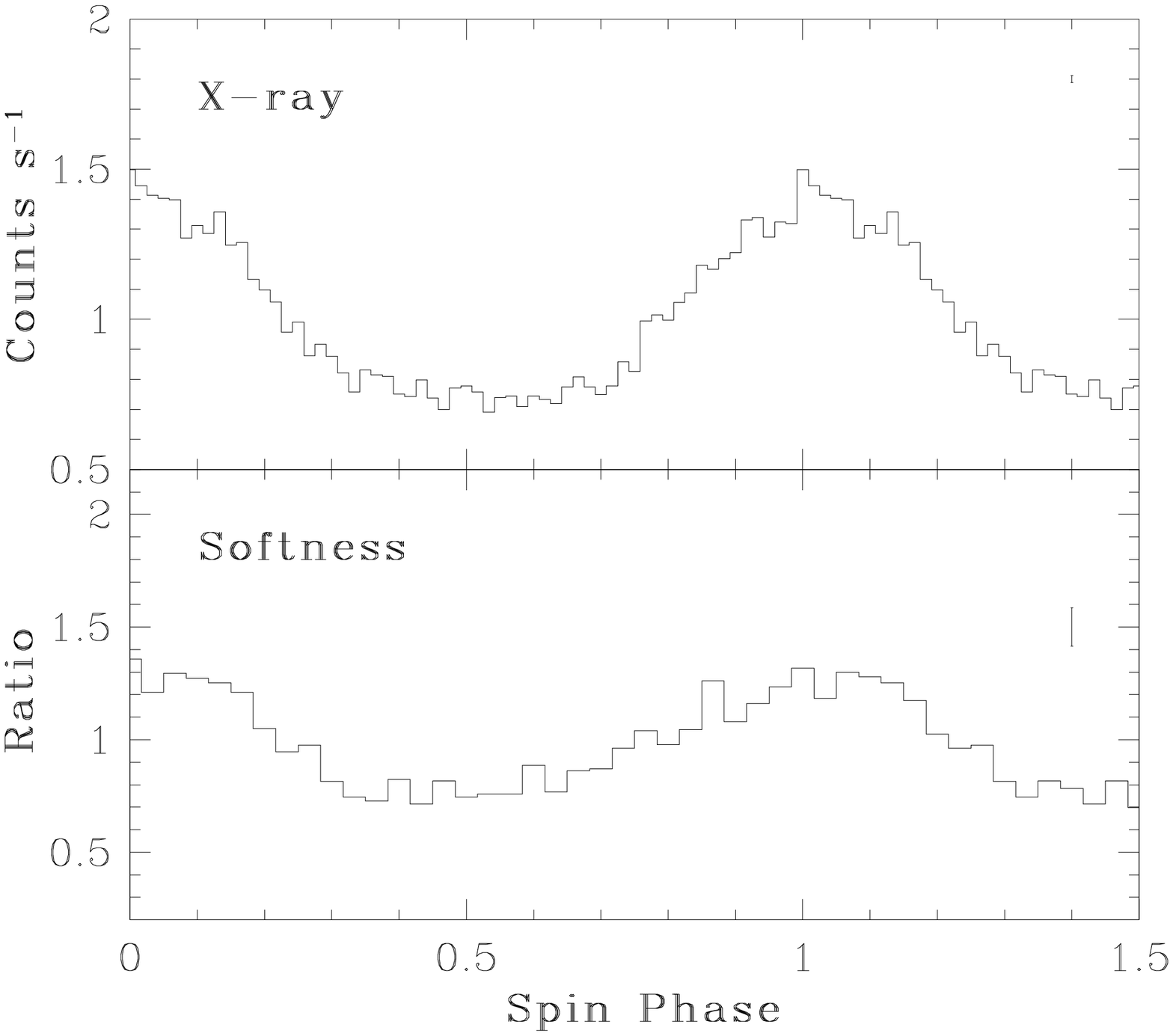}{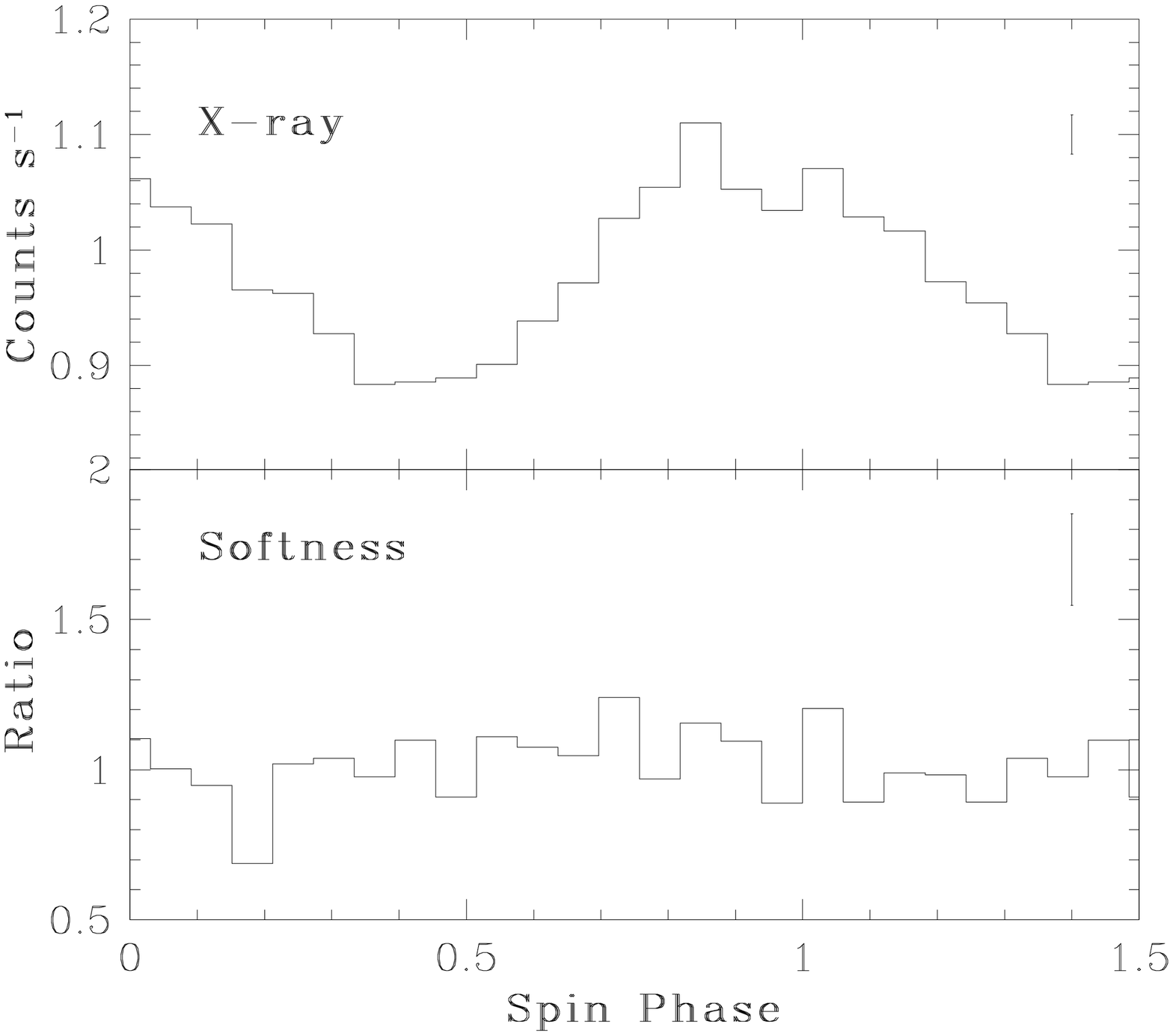}
\caption{As Fig.~1, but for EX~Hya (left) and AE~Aqr (right).}
\label{fig:ex}
\label{fig:ae}
\end{center}
\end{figure}

EX~Hya's softness ratio has the same shape and phasing as the
lightcurve (Fig.~\ref{fig:ex}), making it tempting to interpret this
modulation as changing absorption in the accretion curtains (indeed,
it was for this system that accretion curtains were first proposed as
the cause of X-ray modulation; Rosen, Mason, \&\ C\'ordova 1988).
However, various X-ray studies (e.g.\ Allan, Hellier, \&\ Beardmore
1998) have suggested that the upper emitting pole is periodically
occulted by the white dwarf, giving rise to the modulation. Since
such occultation will affect the lower, cooler parts of the accretion
column more than the higher, hotter ones, we see a deeper modulation
at lower energies.

\section{AE Aquarii}
\label{sec:ae}

AE~Aqr differs from many IPs as it is a rapid rotator ($P_{\rm spin}
\sim$ 33 s), and thus its magnetic field might be expelling material
from the system like a propeller (e.g., Eracleous \&\ Horne 1996). It
is unclear whether the X-ray emission arises from accretion, as is
usual in IPs, or further out in the magnetosphere, as suggested by
Ikhsanov (2001). However, the \emph{XMM-Newton\/} data shows that the
pulse profile is sinusoidal and largely independent of energy.

\section{GK~Perseii (in outburst)}
\label{sec:gkper}

\begin{figure}
\begin{center}
\plottwo{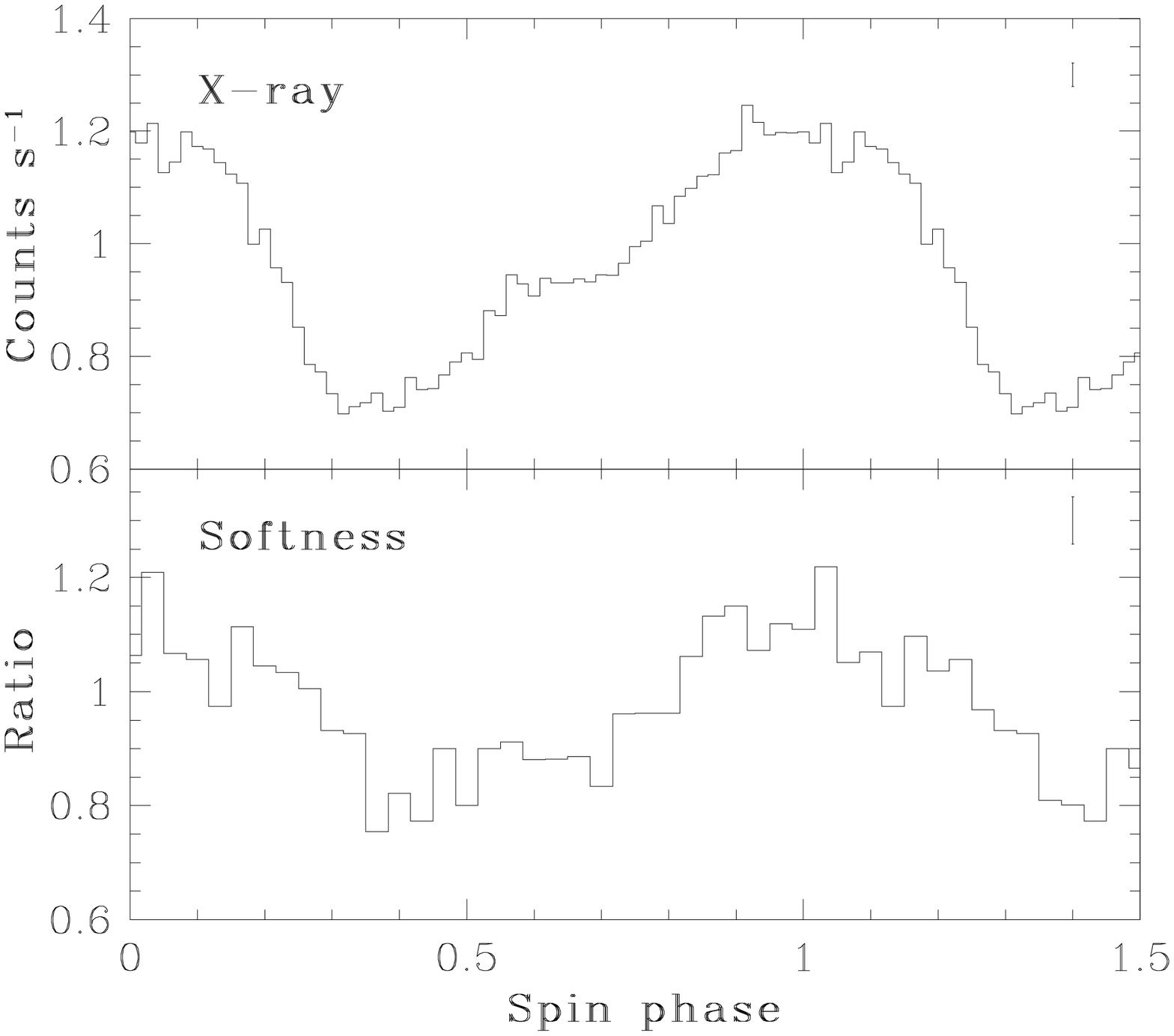}{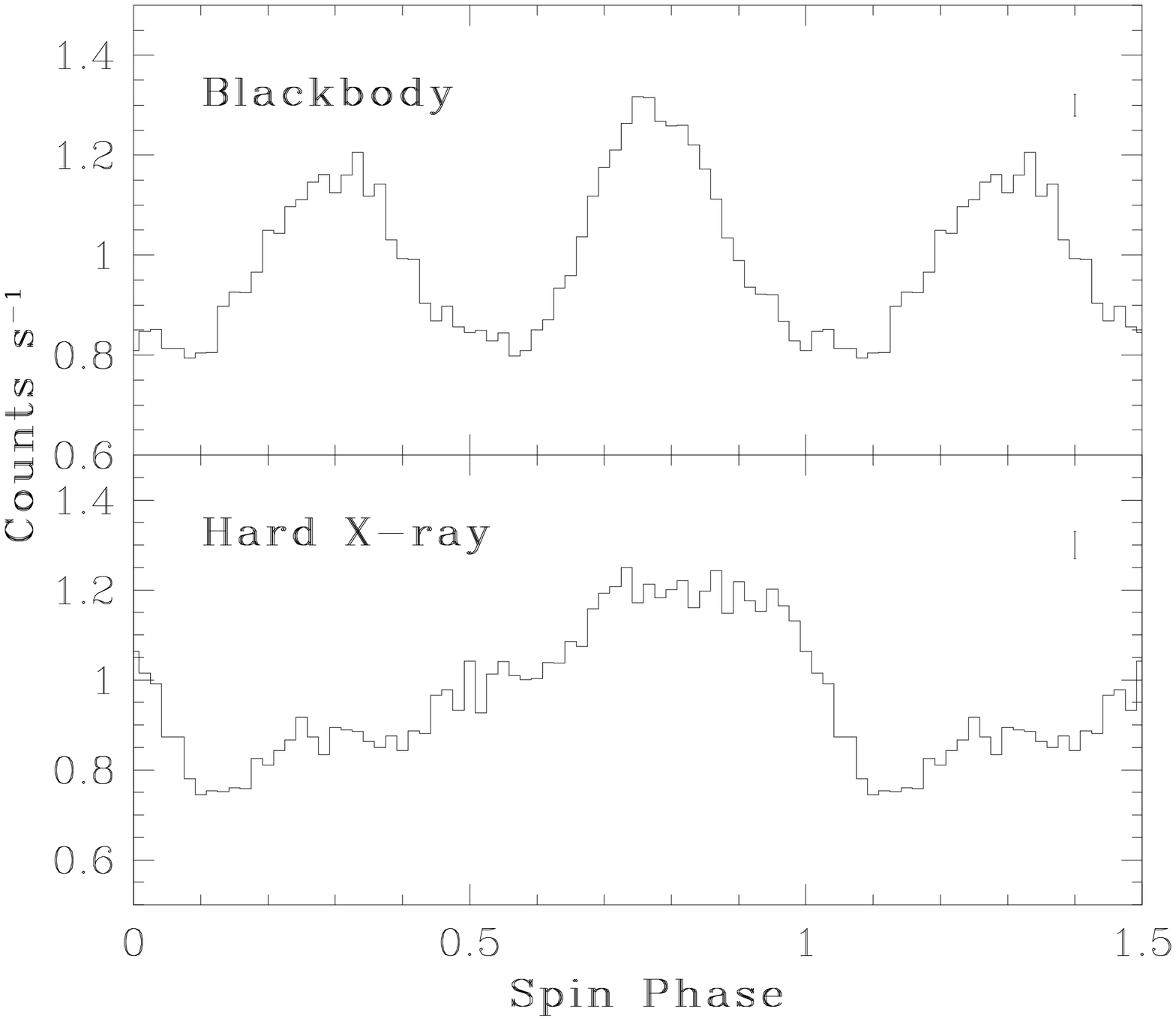}
\caption{Left: As for Fig.~1, but for GK~Per. Right: Spin-pulse
profile of V405~Aur. The upper panel shows the 0.2--0.7 keV band and
the lower panel shows the 0.7--12 keV band.}
\label{fig:v405}
\label{fig:gk}
\end{center}
\end{figure}

Watson, King, \&\ Osborne (1985) suggested that the X-ray spin-pulse
of GK~Per in outburst was caused by increased absorption at spin
minimum. Ishida et al.\ (1992) blamed changing absorption for the
profile in quiescence as well, but suggested that the outburst
profile may show evidence for occultation of the upper pole.
Fig.~\ref{fig:gk} shows that the outburst softness ratio follows the
lightcurve, making it more likely that absorption changes are
responsible for this pulse, as Watson et al.\ (1985) claimed.
Hellier, Harmer, \&\ Beardmore (2004) agree with this, in analysis of
an \emph{RXTE\/} observation.

\section{V405 Aurigae}
\label{sec:v405}

V405~Aur differs from IPs such as AO~Psc, first because it shows a
soft blackbody component to its X-ray emission (Haberl et al.\ 1994),
and second because the blackbody is double-peaked on the spin period,
while the harder emission is single-peaked but sawtoothed
(Fig.~\ref{fig:v405}). Furthermore, de~Martino et al.\ (2004a) and
Evans \&\ Hellier (2004) have shown that the absorption does not
change with spin phase. Evans \&\ Hellier (2004) suggest that if the
angle between the magnetic and spin axes in this system is very high,
then the double-peaked profile of the blackbody emission is
explained.

\section{FO Aquarii}
\label{sec:fo}

\begin{figure}
\begin{center}
\plottwo{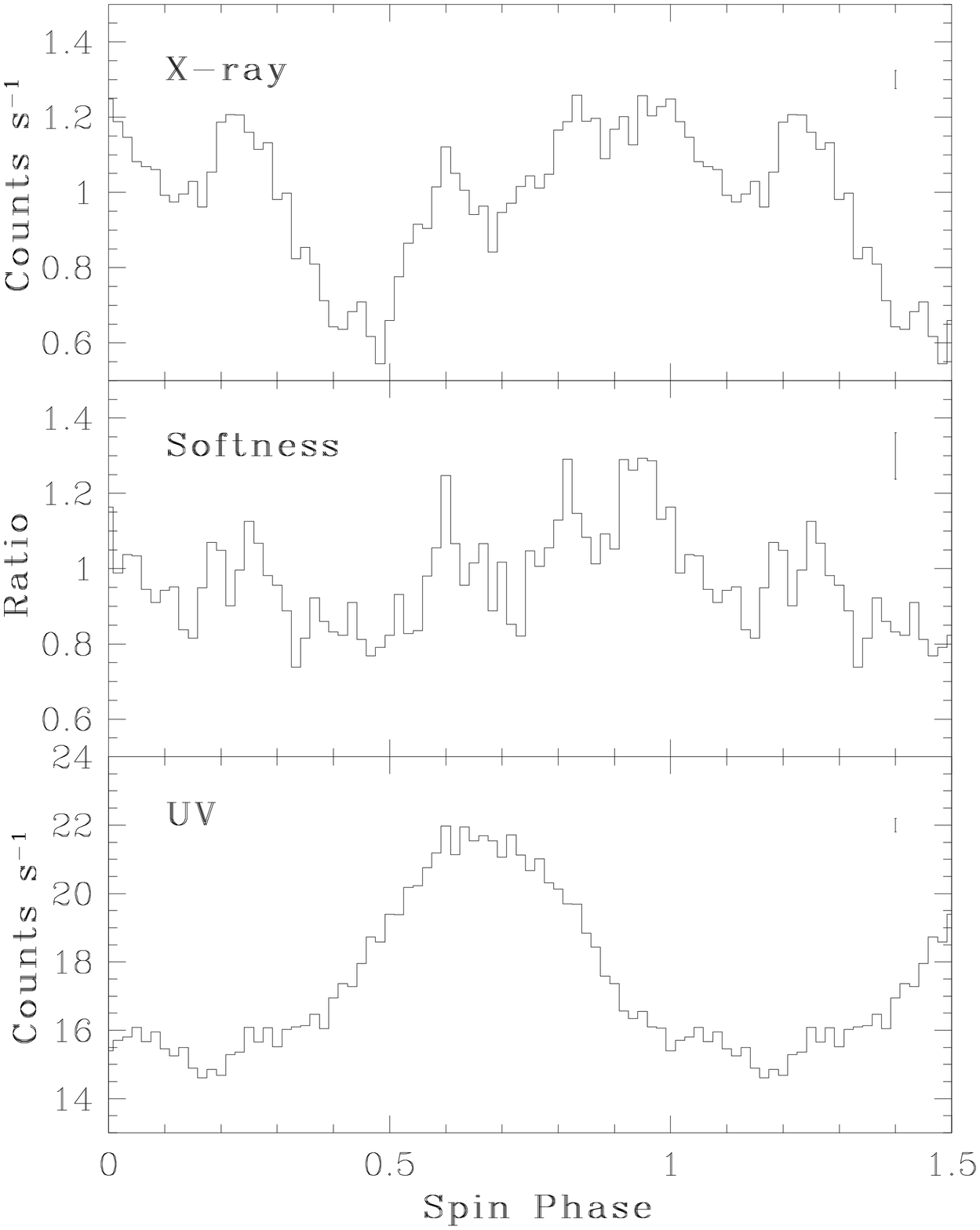}{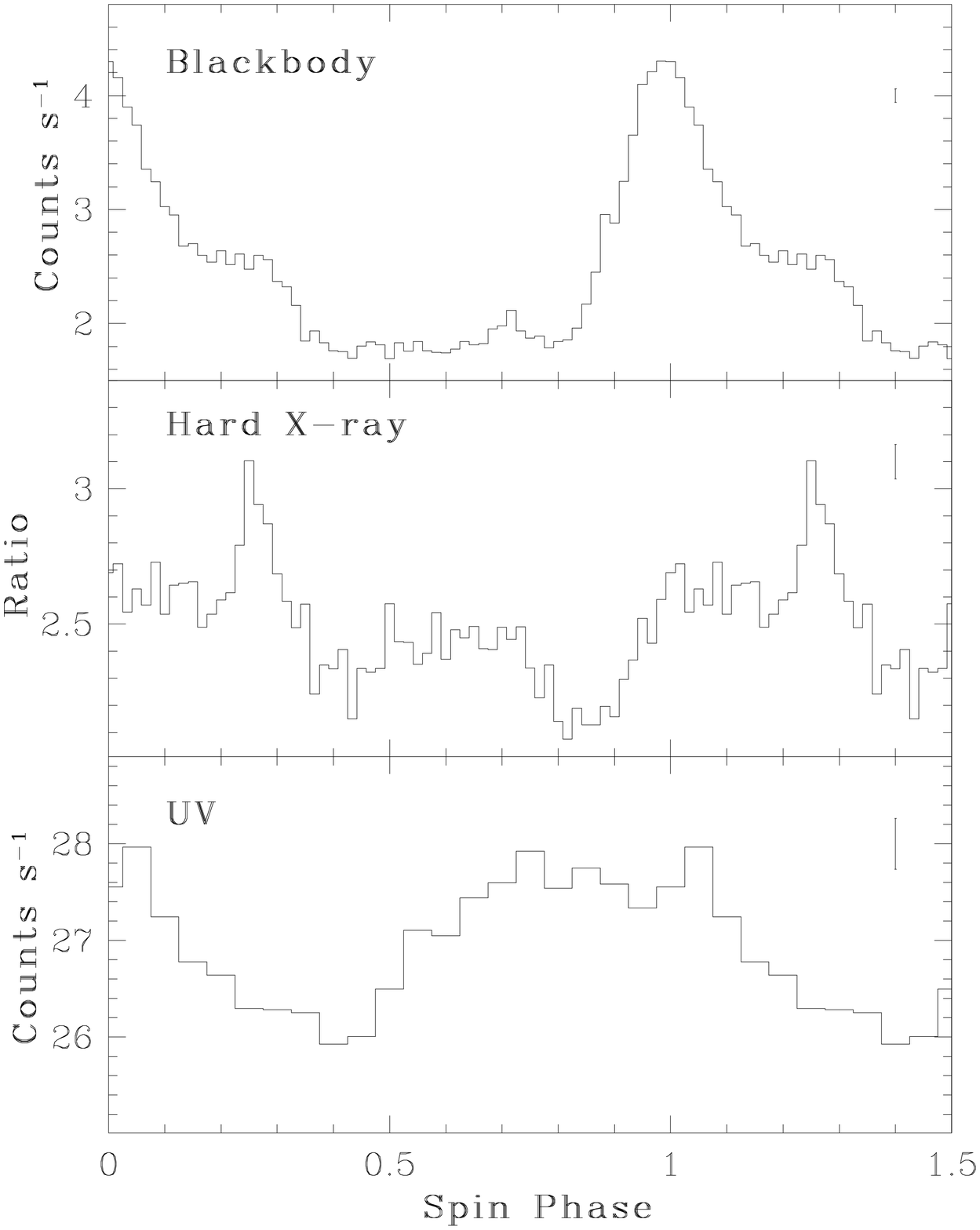}
\caption{Left: as Fig.~1, but for FO~Aqr. Right:  PQ~Gem spin folds
for the blackbody (upper panel), hard X-ray (lower panel) and UV 
(2450--3200 \AA) components.}
\label{fig:fo}
\label{fig:pq}
\end{center}
\end{figure}

The X-ray pulse profile of FO~Aqr (Fig.~\ref{fig:fo}) shows X-ray
minimum to occur after UV minimum. Many authors (e.g.\ Hellier 1993;
Beardmore et al.\ 1998) have identified the `notch' at phase 0.7 as
the result of occultation of the upper accretion column, and the dip
around phase 0.5 as arising from the accretion curtains intercepting
our line of sight. Evans et al.\ (2004) note that, if this is the
case, the upper pole will be pointed towards the observer nearly a
quarter of a cycle before the accretion curtain dip. They thus
suggest that the accretion curtains are twisted, explaining the
lightcurve and softness ratio.

\section{PQ Geminorum}
\label{sec:pq}

PQ~Gem shows a soft blackbody component as well as the hard,
optically thin emission characteristic of IPs (Mason et al.\ 1992).
The spin-pulse profile (Fig.~\ref{fig:pq}) differs greatly from that
of AO~Psc, with UV maximum coinciding with X-ray minimum. Potter et
al.\ (1997) and  Mason (1997) have suggested that the blackbody
modulation is caused by changing views of the accretion region as the
white dwarf rotates. Maximum (phase 0) occurs when the upper pole is
towards us. Since this occurs after the dip around phase 0.8,
interpreted as absorption by the accretion curtains, Potter et al.\
(1997) and Mason (1997) suggest that PQ~Gem accretes preferentially
along field lines preceding the magnetic pole.

\section{V2400 Ophiuchi}
\label{sec:v2400}

\begin{figure}
\begin{center}
\plotfiddle{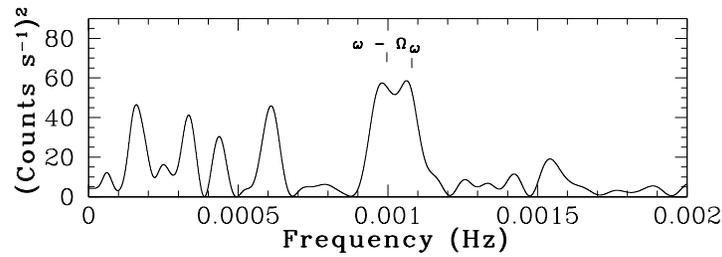}{3cm}{-90}{35}{35}{-130}{200}
\caption{Power spectrum of V2400~Oph. The spin ($\omega$) and beat
($\omega-\Omega$) frequencies are marked.}
\label{fig:v2400}
\end{center}
\end{figure}

V2400~Oph is thought to be the only discless IP, since it shows a
dominant X-ray modulation on the spin-orbit beat period (Buckley et
al.\ 1995; Hellier \&\ Beardmore 2002). We have recently discovered
evidence for a spin pulse during an \emph{XMM-Newton\/} observation
(Fig.~\ref{fig:v2400}), the first in the X-ray band.

\end{document}